\documentclass{appolb}

\usepackage{graphicx}
\usepackage{subfigure}
\usepackage{amssymb}
\usepackage{amsmath}
\usepackage{booktabs}
\usepackage{multirow}
\usepackage{wasysym}
\usepackage{setspace}

\begin{document}
\title{BFKL Amplitude Parametrisation for the Jet-Gap-Jet Events at the LHC Energies
\thanks{Talk presented by Paula \'Swierska at 58$^{\mathrm{th}}$ Scientific Conference for Students of Physics and Natural Sciences, 24-27 March 2015, Vilnius, Lithuania.}%
}
\author{Paula \'Swierska
\address{Faculty of Physics, Mathematics and Computer Science, Cracow University of Technology, Poland}
\and
Maciej Trzebi\'nski\thanks{Corresponding author; e-mail: maciej.trzebinski@ifj.edu.pl}
\address{H. Niewodnicza\'nski Institute of Nuclear Physics Polish Academy of Sciences\\
ul. Radzikowskiego 152, 31-342 Krak\'ow, Poland.}
}
\maketitle

\begin{abstract}
The process of jet-gap-jet (JGJ) production is briefly described. The JGJ scattering amplitude parametrisation is discussed. On the basis of full amplitude calculations, the parametrisation formulas for the leading logarithmic (LL) and next-to-leading logarithmic (NLL) approximations are obtained. For each case a sum over all conformal spins is considered. The obtained agreement is better than 0.25\% for LL and 1\% for NLL.
\end{abstract}


\section{Introduction}

Hard diffractive processes have been an important part of the studies performed in high energy physics since their discovery in the UA8 experiment~\cite{UA8}. The data collected by HERA and Tevatron detectors allowed to deepen these studies. Nevertheless, now, in the LHC era, many questions still remain open.

The definition of diffraction is usually connected with an exchange of a colourless object. In the case of the electromagnetic fields the exchange is mediated by a photon, whereas in the case of strong interactions it is related to the Pomeron. A colourless exchange leads to one of the most prominent features of diffraction -- the presence of a large rapidity gap. In theory, a gap is a space interval in rapidity in which no particles are produced. Moreover, since the colourless exchange does not influence the quantum numbers, the state of interacting objects is preserved.

\begin{figure}[!htbp]
\centering
\includegraphics[width=0.4\textwidth]{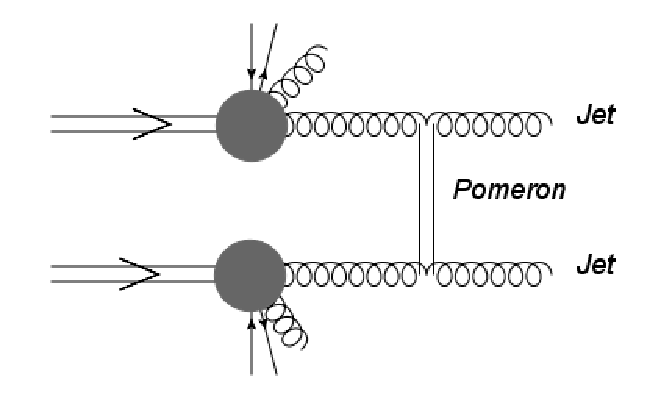}
\caption{Diagram of the jet-gap-jet production.}
\label{fig:JGJ_diag}
\end{figure}

In particular, jet events could be created as a result of the colourless exchange between interacting gluons (see Fig.~\ref{fig:JGJ_diag}). Such a process will be hereafter called the jet-gap-jet (JGJ) production.

The presence of a hard scale in the jet-gap-jet events makes it possible to understand these events with perturbative methods. However, after many theoretical investigations, there is still no consensus on what is the relevant QCD mechanism. When the rapidity gap is sufficiently large, the perturbative QCD description of the jet-gap-jet events is usually performed in terms of the Balitsky-Fadin-Kuraev-Lipatov (BFKL) Pomeron~\cite{BFKL}. As was shown in Ref.~\cite{Royon_gap}, this model can be tested experimentally by studying the behaviour of the ratio of dijet events with a gap to all dijet events as a function of the leading jet transverse momentum or the interval in rapidity between the two leading jets.

It is worth mentioning that the jet-gap-jet events can also be produced in single diffractive and double Pomeron exchange topologies~\cite{DPE_JGJ}. Such events, so far never observed experimentally, are expected to be measured at the LHC.

In order to account for the experimental effects, the detector simulation is usually performed. The events used in such simulations should have properties similar to the ones that are to be measured. This requires a use of the Monte Carlo (MC) generators. The process of the jet-gap-jet production was implemented in several such tools, \textit{e.g.} \textsc{Herwig}~\cite{Herwig} or FPMC~\cite{FPMC}. In order to speed up the calculations, the cross-section formulas were parametrised~\cite{Royon_gap}. These equations were fitted assuming that the transverse momentum of the leading jet did not exceed 120~GeV. However, as this assumption is no longer valid at the energies available at the LHC, a new parametrisation has to be carried out. In this paper we present new fit formulas applicable for the LHC energies.

\section{Jet-Gap-Jet Formalism}
The parton-level cross-section for the jet-gap-jet production can be calculated as~\cite{Royon_gap}:
$$ \frac{d^3 \sigma^{pp\rightarrow XJJY}}{dx_{1}dx_{2}dp^{2}_{T}} = S^2 \cdot f_{eff}(x_1,p^2_T) \cdot f_{eff}(x_2,p_T^2) \frac{d\sigma^{gg\rightarrow gg}}{dp^2_T},$$
where $S^2$ is the gap survival probability, $f_{eff}(x_{1,2},p^2_T)$ are the effective Parton Distribution Functions (PDFs):
$$f_{eff}(x_{1,2},p^2_T) = g(x, p_T^2) + \frac{C_F^2}{N_c^2}\left( q(x, p_T^2) + \bar{q}(x, p_T^2) \right),$$
where $N_C = 3$ is the number of colours, $C_F$ -- the QCD colour factor and $g$, $q$ and $\bar{q}$ are the gluon, quark, antiquark distribution functions in the interacting hadrons. The $\frac{d\sigma^{gg\rightarrow gg}}{dp^2_T}$ cross-section is given by:
\begin{equation}
\label{eq_amplitude}
\begin{split}
\frac{d\sigma^{gg\rightarrow gg}}{dp^2_T} & = \frac{1}{16\pi} |A(\Delta\eta, p^2_T)|^2 \\
& = \frac{16 N_c^2 \pi \alpha^4_s}{C_F^2 p^4_T} \left| \sum_{p=-\infty}^{\infty}\int \frac{d\gamma}{2i\pi}
\frac
{[p^2-(\gamma - \frac{1}{2})^2] \cdot \exp\{\bar{\alpha}(p^2_T) \cdot \chi_{eff} \cdot \Delta\eta\}}
{[(\gamma-\frac{1}{2})^2-(p-\frac{1}{2})^2][(\gamma-\frac{1}{2})^2-(p+\frac{1}{2})^2]} \right|^2,
\end{split}
\end{equation}
where $\chi_{eff}$ -- the effective BFKL kernel~\cite{Royon_gap}, $\alpha^2_s(p_T^2) = \pi \bar{\alpha}^2_s(p_T^2)/N_C$ is the running coupling constant and $p$ denotes the conformal spin. The complex integral runs along an imaginary axis from $\frac{1}{2} - i\infty$ to $\frac{1}{2} + i\infty$. The necessity of summing up all the conformal spins was demonstrated in~\cite{conformal_spin}. The BFKL kernel was calculated in the leading logarithmic (LL)~\cite{BFKL} and next-to-leading logarithmic (NLL)~\cite{BFKL_NLL} approximations.

\section{Jet-Gap-Jet Amplitude Parametrisation}

In principle, the JGJ amplitude can be directly implemented into the Monte Carlo generator. Unfortunately, due to the complexity of Eq.~\ref{eq_amplitude}, the computation time is quite long. Since the usual number of events for a physics analysis is of the order of $10^5$, the parametrisation procedure was postulated to speed up the generation process~\cite{Royon_gap}. This parametrisation was done for the phase-space expected to be measurable using the Tevatron data~\cite{Tevatron_data}:
\begin{itemize}
  \item transverse momentum of jets: $20 < p_T < 120$ GeV,
  \item pseudorapidity distance between jets: $0 < \Delta\eta < 10$.
\end{itemize}
\noindent The conformal spins were summed from $p = -10$ to $p = 10$, which is sufficient for the considered jet $p_T$ range~\cite{Royon_gap}. The parametrisation procedure resulted in the speed-up factors of 10 and 1000 for LL and NLL, respectively. In the following, we redo the parametrisation for the values expected to be observed at the LHC~\cite{LHC_veto}: the range of jet transverse momentum extended to 1 TeV and the sum over conformal spins from -50 to 50.

\subsection{Leading Logarithmic Approximation}
The leading logarithmic approximation is known to be insufficient as the next-to-leading BFKL corrections are expected to be large~\cite{BFKL_NLL}. However, for the completeness, we discuss it below. Denoting $z(p_T^2) = \bar{\alpha}^2_s(p_T^2) \cdot \Delta\eta/2$, the LL cross-section can be parametrised as:
\begin{itemize}
  \item $A_{LL}^{p=0}(z) = N \cdot \left[ A + \exp(B + C \cdot z + D \cdot z^2 + E \cdot z^3+ F \cdot z^4) \right]$, for $p=0$,
  \item $A_{LL}^{all\ p}(z) = N \cdot \left[ A + B \cdot z + \exp(C + D \cdot z + E \cdot z^2 + F \cdot z^3) \right]$, for sum over conformal spins,
\end{itemize}
where the normalisation constant is equal to $N = \frac{\bar{\alpha}^2_s}{4\pi}$ with $\alpha^2_s$ fixed to 0.17.

\begin{figure}[!htbp]
\centering
\includegraphics[width=0.49\textwidth]{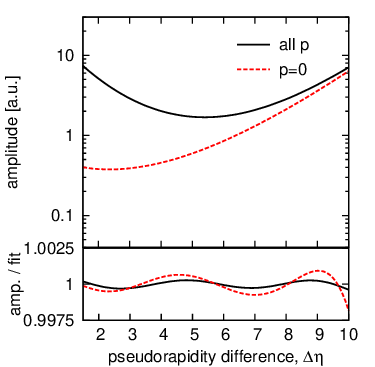}
\caption{\textbf{Top:} the shape of the full amplitude as a function of pseudorapidity difference. \textbf{Bottom:} comparison of the leading logarithmic full amplitude calculations and parametrisation results (fit). Black, solid line is for the sum over all conformal spins and the red, dashed one is for $p=0$.}
\label{fig:LL_fit}
\end{figure}

The shape of the full amplitude as a function of pseudorapidity difference is shown in Fig.~\ref{fig:LL_fit} (top). In the bottom, the comparison between the full amplitude calculations and the parametrisation results (fit) is presented. The obtained fit parameters are listed in Table~\ref{tab:LL_fit}. For both considered cases the differences are well below $2.5\permil$ for the whole pseudorapidity range.

\begin{table}[!htbp]
  \caption{Fit parameters for leading logarithmic amplitude parametrisation for $p=0$ and sum over conformal spins (all p).}
  \label{tab:LL_fit}
  \begin{center}
  {\small
    \begin{tabular}{c | r@{\hskip 1mm} c@{\hskip 1mm} l | r@{\hskip 1mm} c@{\hskip 1mm} l}
    \toprule
\textbf{Parameter} & \multicolumn{3}{c|}{$\mathbf{p=0}$} & \multicolumn{3}{c}{\textbf{all p}} \\
    \midrule
A & $0.452$&$\pm$&$0.023$   & $-2.032$&$\pm$&$ 0.022$\\
B & $2.2162$&$\pm$&$0.0022$ & $-1.135$&$\pm$&$0.057$\\
C & $-6.436$&$\pm$&$0.029$  & $6.18035$&$\pm$&$0.00012$\\
D & $21.16$&$\pm$&$0.12$    & $-14.3093$&$\pm$&$0.0032$\\
E & $-16.46$&$\pm$&$0.14$   & $20.650$&$\pm$&$0.011$\\
F & $5.586$&$\pm$&$0.056$   & $-6.4983$&$\pm$&$0.0096$\\
    \bottomrule
  \end{tabular}
  }
  \end{center}
\end{table}

\subsection{Next-to-Leading Logarithmic Approximation}
\begin{figure}[!htbp]
\centering
\includegraphics[width=0.49\textwidth]{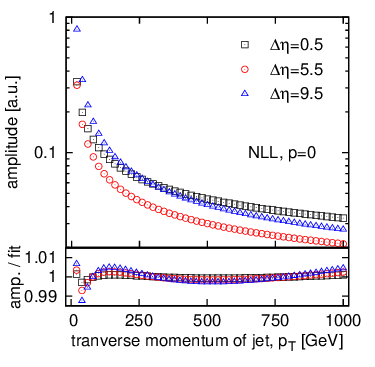}\hfill
\includegraphics[width=0.49\textwidth]{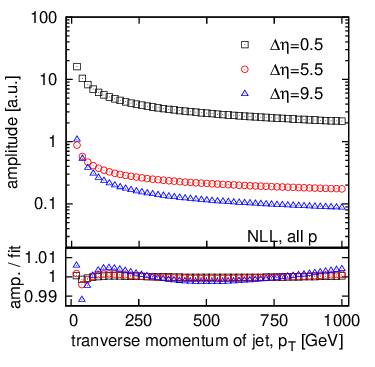}
\caption{\textbf{Top:} the shape of the full amplitude as a function of jet transverse momentum. \textbf{Bottom:} comparison between next-to-leading logarithmic full amplitude calculations and parametrisation results (fit) for $p=0$ (\textbf{left}) and sum over all conformal spins (\textbf{right}). Black rectangles are for $\Delta\eta = 0.5$, red circles for $\Delta\eta = 5.5$ and blue triangles for $\Delta\eta = 9.5$.}
\label{fig:NLL_fit_pt}
\end{figure}

The parametrisation of NLL amplitude is more complex as it depends on both: jets transverse momenta (identical for both jets in collinear approach) and their pseudarapidity distance. In order to obtain the parametrisation formulas, the one-dimensional problem of finding the general dependence on the jet transverse momentum was addressed first. This formula was found to be:
\begin{equation}
A_{NLL}(p_T, \Delta \eta = \mathrm{fixed}) = N \cdot \left[ A(\Delta \eta) \cdot p_T^{B(\Delta \eta)} + C(\Delta \eta) \cdot p_T^{D(\Delta \eta)} \right].
\label{eq:NLL_amp}
\end{equation}
The exemplary results obtained for three different pseudorapidity distances ($\Delta\eta = 0.5$, $\Delta\eta = 5.5$ and $\Delta\eta = 9.5$) are shown in Fig.~\ref{fig:NLL_fit_pt}. In the top part of this figure the shape of the distribution is presented, whereas in the bottom the ratios of full amplitude calculation to fit are shown. The overall agreement is well below 1\%, which means that Eq.~\ref{eq:NLL_amp} works for the whole considered $\Delta \eta$ range.

The parametrisation formulas for both cases ($p=0$ and the sum over all conformal spins) were found to have the following forms:
\begin{itemize}
  \item $A(z) = a_0 + a_1 \cdot z + \exp(a_2 + a_3 \cdot z + a_4 \cdot z^2 + a_5 \cdot z^3)$,
  \item $B(z) = b_0 + b_1 \cdot z$,
  \item $C(z) = c_0 + c_1 \cdot z + \exp(c_2 + c_3 \cdot z + c_4 \cdot z^2 + c_5 \cdot z^3)$,
  \item $D(z) = d_0 + d_1 \cdot z + d_2 \cdot z^2 + d_3 \cdot z^3$.
\end{itemize}

\begin{figure}[!htbp]
\centering
\includegraphics[width=0.49\textwidth]{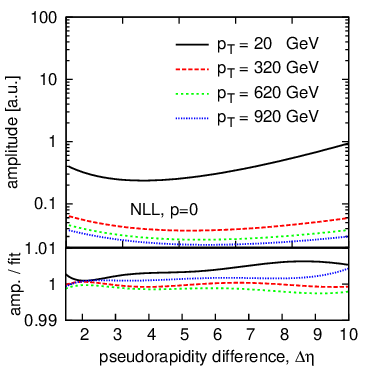}\hfill
\includegraphics[width=0.49\textwidth]{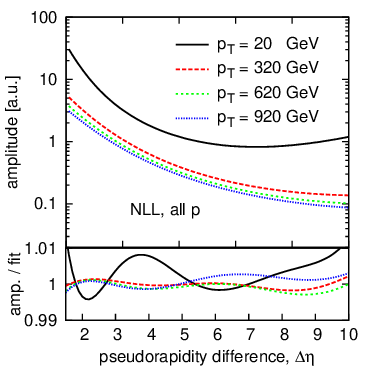}
\caption{\textbf{Top:} the shape of the full amplitude as a function of jet transverse momentum. \textbf{Bottom:} Comparison between next-to-leading logarithmic full amplitude calculations and parametrisation results (fit) for $p=0$ (\textbf{left}) and sum over all conformal spins (\textbf{right}) as a function of presudorapidity difference. Solid black lines are for jet transverse momentum of $p_T = 20$~GeV, red dashed ones for $p_T = 320$~GeV, green fine-dashed for $p_T = 620$~GeV and blue dotted for $p_T = 920$~GeV.}
\label{fig:NLL_fit}
\end{figure}

The shape of the full amplitude as a function of pseudorapidity difference for the NLL is shown in Fig.~\ref{fig:NLL_fit} (top). In the bottom, the comparison of the full amplitude calculations and the parametrisation results is shown. The obtained fit parameters are listed in Table~\ref{tab:NLL_fit}. For both considered cases the differences are below 1\% for the whole considered pseudorapidity range.

\begin{table}[!htbp]
  \caption{Fit parameters for next-to-leading logarithmic amplitude
parametrisation for $p=0$ and sum over conformal spins (all p).}
  \label{tab:NLL_fit}
  \begin{center}
{\small
    \begin{tabular}{c | r@{\hskip 1mm} c@{\hskip 1mm} l | r@{\hskip 1mm} c@{\hskip 1mm} l}
    \toprule
\textbf{Parameter} & \multicolumn{3}{c|}{$\mathbf{p=0}$} & \multicolumn{3}{c}{\textbf{all p}} \\
\midrule
$a_0$ & $-51.3$ & $\pm$ & $2.0$   & $-24.7$ & $\pm$ & $1.8$\\
$a_1$ & $28.3$ & $\pm$ & $9.5$   & $235.6$ & $\pm$ & $2.5$\\
$a_2$ & $4.755$ & $\pm$ & $0.019$   & $7.606$ & $\pm$ & $0.019$\\
$a_3$ & $-2.13$ & $\pm$ & $0.23$   & $-19.97$ & $\pm$ & $0.21$\\
$a_4$ & $8.76$ & $\pm$ & $0.52$   & $36.60$ & $\pm$ & $0.45$\\
$a_5$ & $-3.14$ & $\pm$ & $0.31$ & $-16.56$ & $\pm$ & $0.29$\\
\midrule
$b_0$ & $-0.9751$ & $\pm$ & $0.0034$   & $-0.6666$ & $\pm$ & $0.0041$\\
$b_1$ & $-0.7556$ & $\pm$ & $0.0058$   & $-0.9422$ & $\pm$ & $0.0064$\\
\midrule
$c_0$ & $-0.58$ & $\pm$ & $0.12$   & $0.826$ & $\pm$ & $0.045$\\
$c_1$ & $0.300$ & $\pm$ & $0.064$   & $1.72$ & $\pm$ & $0.12$\\
$c_2$ & $2.031$ & $\pm$ & $0.018$   & $6.194$ & $\pm$ & $0.014$\\
$c_3$ & $-2.97$ & $\pm$ & $0.13$ & $-14.564$ & $\pm$ & $0.088$\\
$c_4$ & $6.87$ & $\pm$ & $0.29$ & $16.51$ & $\pm$ & $0.21$\\
$c_5$ & $-2.23$ & $\pm$ & $0.18$ & $-5.22$ & $\pm$ & $0.15$\\
\midrule
$d_0$ & $-0.3880$ &$\pm$& $0.0018$   & $-0.3681$ &$\pm$& $0.0015$\\
$d_1$ & $0.096$ &$\pm$& $0.011$   & $0.7878$ &$\pm$& $0.0093$\\
$d_2$ & $-0.547$ &$\pm$& $0.025$   & $-1.423$ &$\pm$& $0.020$\\
$d_3$ & $0.216$ &$\pm$& $0.017$   & $0.586$ &$\pm$& $0.014$\\
    \bottomrule
  \end{tabular}
}
  \end{center}
\end{table}

\section{Summary}

The process of jet-gap-jet production is very interesting to be studied. The investigation of its properties would allow us not only to measure the cross-section, but also to test the BFKL model.

The previous parametrisation, constructed to describe the Tevatron data, was extended to include phase-space regions available at the LHC energies. The jet transverse momentum was assumed to be within the 20~GeV to 1~TeV range and the conformal spins were summed from -50 to 50. The obtained agreement between the full amplitude calculations and the parametrisation results was found to be better than 0.25\% for leading logarithmic and 1\% for next-to-leading logarithmic approximations.

\section{Acknowledgements}
We gratefully acknowledge Christophe Royon for providing his code for the JGJ full amplitude calculation. We thank Janusz Chwastowski and Rafa{\l} Staszewski for discussions and suggestions. This work was partially supported by the Polish National Science Centre grants number UMO-2012/05/B/ST2/02480 and UMO-2012/05/N/ST2/02697.


\begin{thebibliography}{99}
\bibitem{UA8} UA8 Collaboration (A. Brandt \textit{et. al}), \textit{Evidence for Transverse Jets in High Mass Diffraction}, Phys. Lett. B \textbf{211} (1988) 239,\\
UA8 Collaboration (A. Brandt \textit{et. al}), \textit{Cross-section measurements of hard diffraction at the SPS collider}, Phys. Lett. B \textbf{421} (1998) 395.
%
\bibitem{BFKL} L. N. Lipatov, \textit{Reggeization of the Vector Meson and the Vacuum Singularity in Non-abelian Gauge Theories}, Sov. J. Nucl. Phys. \textbf{23} (1976) 338,\\
E. A Kuraev, L. N. Lipatov and V. S. Fadin, \textit{The Pomeranchuk Singularity in Non-abelian Gauge Theories}, Sov. Phys. JETP \textbf{45} (1977) 199,\\
I. I. Balitsky and L. N. Lipatov, \textit{The Pomeranchuk Singularity In Quantum Chromodynamics}, Sov. J. Nucl. Phys. \textbf{28} (1978) 822.
%
\bibitem{Royon_gap} F. Chevallier, O. Kepka, C. Marquet, C. Royon, \textit{Gaps between jets at hadron colliders in the next-to-leading BFKL framework}, Phys. Rev. D \textbf{79} (2009) 094019,\\
O. Kepka, C. Marquet, C. Royon, \textit{Gaps between jets in hadronic collisions}, Phys. Rev. D \textbf{83} (2011) 034036.
%
\bibitem{DPE_JGJ} C. Marquet, C. Royon, M. Trzebinski, R. Zlebcik, \textit{Gaps between jets in double-Pomeron-exchange processes at the LHC}, Phys. Rev. D \textbf{87} (2013) 034010.
%
\bibitem{Herwig} G. Corcella \textit{et al.}, \textit{HERWIG 6.5: an event generator for Hadron Emission Reactions With Interfering Gluons}, JHEP \textbf{0101} (2001) 010.
%
\bibitem{FPMC} M. Boonekamp \textit{et al.}, \textit{FPMC: a generator for forward physics}, \verb+fpmc.web.cern.ch/project-fpmc/+.
%
\bibitem{conformal_spin} L. Motyka, A. D. Martin and M. G. Ryskin, \textit{The non-forward BFKL amplitude and rapidity gap physics}, Phys. Lett. B \textbf{524} (2002) 107,\\
C. Marquet and C. Royon, \textit{Azimuthal decorrelation of Mueller-Navelet jets at the Tevatron and the LHC}, Phys. Rev. D \textbf{79} (2009) 034028.
%
\bibitem{BFKL_NLL} V. S. Fadin and L. N. Lipatov, \textit{BFKL pomeron in the next-to-leading approximation}, Phys. Lett. B \textbf{429} (1998) 127,\\
M. Ciafaloni and G. Camici, \textit{Energy Scale(s) and Next-to-leading BFKL Equation}, Phys. Lett. B \textbf{430} (1998) 349.
%
\bibitem{Tevatron_data} CDF Collaboration (B. Abbott \textit{et al.}), \textit{Probing hard color-singlet exchange in $p\bar{p}$ collisions at $\sqrt{s}$ = 630 GeV and 1800 GeV}, Phys. Lett. B \textbf{440} (1998) 189,\\
CDF Collaboration (F. Abe \textit{et al.}), \textit{Dijet Production by Color-Singlet Exchange at the Fermilab Tevatron}, Phys. Rev. Lett. \textbf{80} (1998) 1156.
%
\bibitem{LHC_veto} ATLAS Collaboration, \textit{Measurement of dijet production with a veto on additional central jet activity in pp collisions at sqrt(s)=7 TeV using the ATLAS detector}, JHEP \textbf{1109} (2011) 053,\\
CMS Collaboration, \textit{Cross section measurement for simultaneous production of a central and a forward jet in proton-proton collisions at sqrt(s)=7 TeV}, CMS-PAS-FWD-10-006.
\end{thebibliography}
\end{document}